\documentclass[twocolumn,showpacs,preprintnumbers,amsmath,amssymb,aps,prb]{revtex4-1}
\usepackage{graphicx}% Include figure files

\usepackage{dcolumn}% Align table columns on decimal point
\usepackage{bm}% bold math
\usepackage{mathrsfs}
\usepackage{epsfig}
\usepackage{graphicx}
\usepackage{amsmath}
\usepackage{amsfonts}
\usepackage{amssymb}
\usepackage{mathrsfs}
\usepackage{multirow}

\begin{document}
%\title{The lattice structure due to the non-monotonic interaction}
\title{Pattern formation and phase transitions in systems with non-monotonic interaction}

\author{H.J. Zhao}
\affiliation{
Department of Physics, University of Antwerpen, Groenenborgerlaan 171, B-2020 Antwerpen,
Belgium
}
\author{V.R. Misko}
\affiliation{
Department of Physics, University of Antwerpen, Groenenborgerlaan 171, B-2020 Antwerpen,
Belgium
}

\author{F.M. Peeters}
% \email{francois.peeters@ua.ac.be}
\affiliation{
Department of Physics, University of Antwerpen, Groenenborgerlaan 171, B-2020 Antwerpen,
Belgium
}

\date{\today}% It is always \today, today,

\begin{abstract}
We analyzed pattern formation and identified different phases in a system
of particles interacting through a non-monotonic short-range repulsive
($r<r_{c}$) and long-range attractive ($r>r_{c}$) potential,
using molecular-dynamics simulations.
Depending on parameters, the interaction potential models the interparticle
interaction in various physical systems ranging from atoms, molecules
and colloids to neutron stars and vortices in low $\kappa$ type-II
superconductors and in recently discovered ``type-1.5'' superconductors.
We constructed the phase diagram ``critical radius $r_{c}$ $-$ density $n$''
and proposed a new approach to characterization of the phases.
Namely, we elaborated a set of quantitative criteria in order to identify
the different phases using the Radial Distribution Function (RDF), the
local density function, and the occupation factor.
\end{abstract}

%\pacs{74.25.Qt,74.25.Ha,74.78.Na}% PACS, the Physics and Astronomy
\maketitle

\section{Introduction}

%Fabricating specific nanometer-sized structures on surfaces is significantly
%important nowadays.
%One attractive approach is to form patterns of nanometer dimensions on solid
%surfaces and then to grow the nanostructures on these templates.

Pattern formation in many systems is governed by competing interaction~\cite{Vedmedenko2007}.
Examples of such systems are: the pasta phase in neutron stars\cite{PhysRevC.69.045804},
ferrofluids\cite{PhysRevE.67.021402,PhysRevE.64.041506,Rosensweig1983127,Tsebers1980}, Langmuir
monolayers\cite{Kaganer1999,Suresh1988}, magnetic garnet thin films,
type-I superconductors~\cite{PhysRevLett.103.267002},
colloids and gels\cite{PhysRevLett.97.078301,Lu2008,PhysRevLett.94.208301,PhysRevLett.104.165702,PhysRevE.77.031608}, etc.
In general, there is a strong correlation between the pattern formation and
the inter-particle interaction.
Attraction favors aggregation, while repulsion favors low local densities.
This competition between repulsive and attractive interaction leads to very rich phases, such as stripes, clusters,
bubbles, etc.\cite{Seul27011995}.
Those different phases were observed in many diverse systems.
In neutron stars, the competition between the short-range nuclear attraction
and the long-range Coulomb repulsion leads to complex pasta phases\cite{PhysRevC.69.045804}.
In ferrofluid systems, rich phases due to the competition between dipolar forces
and short-range forces opposing density variations were found
experimentally\cite{PhysRevE.67.021402} and theoretically\cite{PhysRevE.64.041506}.

Pattern formation was extensively studied in colloidal systems where
the colloid-colloid interaction is characterized by the competition of the
hardcore excluded volume interaction, on the one hand, and the polarization
of the particles, on the other hand.
The interaction potential can be further controlled by adding other components.
As a result, rich configurations, such as clusters, repulsive or attractive glassy
states, gels, were found numerically \cite{Reichhardt2003,PhysRevLett.90.026401,Reichhardt2005,Sciortino2005,PhysRevLett.93.055701,PhysRevB.83.014501}, analytically \cite{PhysRevLett.96.075702,PhysRevE.75.011410} and
experimentally \cite{PhysRevLett.104.165702,Lu2008} in colloidal systems with short range attractive
interaction.
The properties of isolated clusters formed by the short-range attractive and
long-range repulsive interaction were studied\cite{Mossa2004}.
By controlling the interaction, the growth of the cluster was shown to change
from nearly spherical to one-dimensional.
The one-dimensional growth of the clusters facilitated the collective packing
into columnar or lamellar phases \cite{Mossa2004}.
The columnar and lamellar phases in three dimension were also studied by
molecular dynamics (MD) simulations \cite{PhysRevE.74.010403}.

In superconductors, the vortex-vortex interaction is usually considered to be
either repulsive (in type-II superconductors where the Ginzburg-Landau parameter
$\kappa$, i.e., the ratio of the magnetic field penetration depth $\lambda$ to
the coherence length $\xi$, $\kappa>1/\sqrt{2}$) or attractive (in type-I
superconductors, where $\kappa<1/\sqrt{2}$ and vortices are unstable)
while vortices do not interact with each other at the so-called ``dual point''
when $\kappa=1/\sqrt{2}$ \cite{PhysRevB.3.3821}.
However, a deeper analysis of the inter-vortex interaction in type-II superconductors
near the dual point revealed an attractive tail~\cite{Brandt1987,Brandt2011}.
This repulsive-attractive inter-vortex interaction was used for the explanation of
unusual patterns in the intermediate state in low-$\kappa$ superconductors (e.g., Nb):
islands of Meissner phase surrounded by vortex phase or vice versa, i.e.,
vortex clusters surrounded by Meissner phase~\cite{Brandt1987,Brandt2011}.

Recent discovery of ``type-1.5'' superconductors \cite{PhysRevLett.102.117001} induced
a new wave of interest (see, e.g., Ref.~\cite{PhysRevB.83.214523,lucia2})
to systems with non-monotonic interactions, due to the fact
that the observed vortex patterns in those superconductors revealed a clear signature
of the repulsive-attractive inter-vortex interaction.
In particular, several properties (e.g., vortex lattice with voids, the nearest-neighbor
distribution) of the observed vortex patterns were explained using a simple model that
involved a non-monotonic inter-vortex interaction based on a more general approach to
multi-order-parameter condensates~\cite{PhysRevB.72.180502}.

In this paper, we consider a model competing interaction potential which is repulsive
for short range and attractive for long range.
This form of the interaction potential could be used as a model for different systems
with non-monotonic inter-particle interation, e.g., atoms or molecules (i.e., the
Lennard-Jones potential), or vortices in two-band superconductors, depending on specific
parameters of the potential.
We study pattern formation in two-dimensional systems, for different interaction
potential profiles, i.e.,
we distinguish
``soft-core'' and ``hard-core'' interactions and analyze
the transitions between different phases.
Based on this, we construct a phase diagram for different interaction parameters
and particle densities.
We propose a new approach to characterize the different phases:
instead of qualitative characterization of the phases (e.g., clusters or labyrinths),
we introduce a number of quantitative criteria to distinguish these.
In particular,
different phases are analyzed in terms of the Radial Distribution Function (RDF) and
additional quantities chracterizing, e.g., the local density of particles in
clusters.

The paper is organized as follows.
In Sec.~II, we describe the model.
The pattern formation for different interaction parameters is discussed in Sec.~III.
In Sec.~IV, we analyze different phases using the RDF and discuss criteria for
identification of different patterns.
The conclusions are given in Sec.~V.

\section{Model}

The inter-particle interaction potential is taken to be of the following form:
\begin{equation}
\label{eq-V_interaction_1}
%V_{ij}=V_0\left(\frac{a}{b}*K_0(b*r_{ij}/r_0)-K_0(r_{ij}/r_0)\right).
V_{ij}=V_0\left(\frac{a}{b}*K_0(b*r_{ij}/\lambda)-K_0(r_{ij}/\lambda)\right).
\end{equation}
Here,
$K_0$ is a first order Bessel function,
$r_{ij}=\mid\textbf{r}_i-\textbf{r}_j\mid$ is the inter-vortex distance,
$V_0$ and $\lambda$ are the units of energy and length, respectively.
[Note that in case of a type-II superconductor appropriate units of
length $\lambda$ and the energy $V_0$, correspondingly, are the magnetic field
penetration depth $\lambda$
and
$
V_{0} = \Phi_{0}^{2} / 8 \pi^{2} \lambda^{2},
$
where $\Phi_{0} = hc/2e$ (see, e.g., Ref.~\cite{PhysRevB.82.184512})].

In dimensionless form, the interaction potential (\ref{eq-V_interaction_1})
reads as
\begin{equation}
\label{eq-V_interaction_1dl}
V^{\prime}_{ij}=\frac{V_{ij}}{V_0}=\frac{a}{b}*K_0(b*r^{\prime}_{ij})-K_0(r^{\prime}_{ij}),
\end{equation}
where the dimensionless length is defined as $r^{\prime}_{ij}=r_{ij}/\lambda$.
Further on, we will omit the primes and use the dimensionless form of the potential
(\ref{eq-V_interaction_1dl}).
The interaction force is then given by
\begin{equation}
\label{eq-interaction_1}
%\textbf{F}_{ij}=-\mathbf{\nabla}V_{ij}=f_0(a*K_1(b*r_{ij}/r_0)-K_1(r_{ij}/r_0))\hat{\textbf{r}}_{ij},
\textbf{F}_{ij}=-\mathbf{\nabla}V_{ij}=(a*K_1(b*r_{ij})-K_1(r_{ij}))\hat{\textbf{r}}_{ij},
\end{equation}
where
$K_1$ is a first order Bessel function,
%$f_0$ is the unit of force,
$\hat{\textbf{r}}_{ij}=(\textbf{r}_i-\textbf{r}_j)/r_{ij}$,
$a$ and $b$ are two positive coefficients.

\begin{figure}
\begin{center}
\includegraphics[width=0.96\columnwidth]{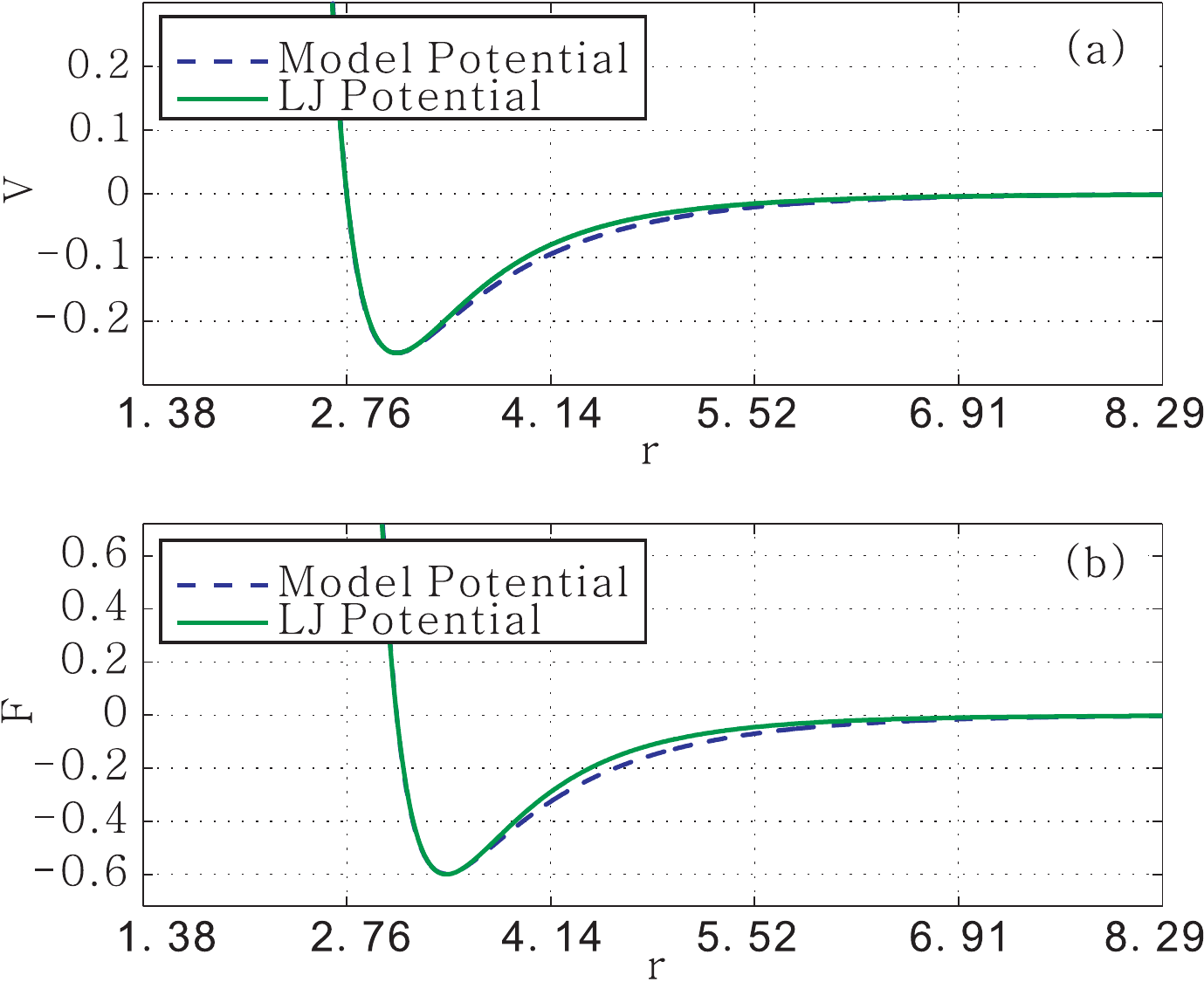}\\
\caption{
(Color online)
The LJ potential and the model potential given by
Eq.~(\ref{eq-V_interaction_1}) with the following parameters:
$a=1.045 \times 10^7$, $b=5.896$, $\sigma=2.762\lambda$, and $\mu_0=0.1V_0$ (a).
The corresponding interaction forces (b).}
\label{fig-Lennard_Jones}
\end{center}
\end{figure}

The interaction potential Eq.~(\ref{eq-V_interaction_1dl}) is generic:
by choosing suitable parameters, it can be used as a model of non-monotonic interactions
in different systems, for example, the well-known inter-atomic (molecule)
Lennard Jones (LJ) potential:
$\mu_{LJ}(r)=\mu_0[(r/\sigma)^{-12}-(r/\sigma)^{-6}]$.
In Fig.~\ref{fig-Lennard_Jones}(a), we compare the LJ potential for
a particle of diameter $\sigma=2.762\lambda$ and the energy unit $\mu_0=0.1V_0$
with the model potential given by Eq.~(\ref{eq-V_interaction_1dl}) with the parameters:
$a=1.045 \times 10^7$, $b=5.896$.
The corresponding interaction forces are presented in Fig.~\ref{fig-Lennard_Jones}(b).
The comparison shows that the potentials and the corresponding forces
(i.e., for the model potential (\ref{eq-V_interaction_1}) and the LJ potential)
fairly agree with each other.
On the other hand, one can easily see that the interaction potential
Eq.~(\ref{eq-V_interaction_1}) is a generalized form of the inter-vortex interaction
in type-II superconductor which, as shown by Kramer\cite{PhysRevB.3.3821}, can be presented
in the form: $V(r)=d_1(\kappa)K_0(r/\lambda)-d_2(\kappa)K_0(r/\xi)$.

The first term of Eq.~(\ref{eq-interaction_1}) is repulsive
while the second term describes an attractive interaction force.
Indeed, for $r\rightarrow\infty$, $K_1(x)\rightarrow\sqrt{\pi/(2x)}e^{-x}$.
The interaction force (\ref{eq-interaction_1}) has a repulsive (attractive) tail
when $b<1$ ($b>1$),
while for $r\rightarrow 0$, $K_1(x)\rightarrow 1/x$
and thus
$\textbf{F}_{ij}\rightarrow (a/b-1)/r$.
Therefore, for short range the interaction force (\ref{eq-interaction_1})
is repulsive (attractive) when $a>b$ ($a<b$),
and we only consider the case $a>b$ since an attractive interaction for
short distances would result in a collapse of our system of point particles.
When $a>b$ and $b<1$, the interaction is always repulsive, and particles
form a Wigner crystal structure.
The most interesting case is realized when $a>b$ and $b>1$, and the interaction
has a repulsive core and attractive tail.
In this case, there exists a critical distance $r_c$, where the inter-particle
interaction energy (\ref{eq-V_interaction_1}) reaches a minimum
(and the interaction force (\ref{eq-interaction_1}) changes sign).
By setting the force equal to zero, the coefficient $a$ is given by
\begin{equation}
\label{eq-coefficent_a}
a=\frac{K_1(r_c)}{K_1(b*r_c)}.
\end{equation}
The pattern formation is determined by the coefficients $b$, $r_c$,
and the particle density $n$.

We study pattern formation in a system of interacting particles using Langevin equations.
The dynamics of a single particle {\it i} obeys the following overdamped equation of motion:
\begin{equation}
\label{Md}
\eta \textbf{v}_i=\textbf{F}_i=\sum_{j\neq i} \textbf{F}_{ij}+\textbf{F}_{i}^T.
\end{equation}
Here, $\eta$ is the viscosity, which is set to unity.
$\textbf{F}_{ij}$ is the interparticle interaction force
defined by Eq.~(\ref{eq-interaction_1})
and
$\textbf{F}_{i}^T$ is the stochastic thermal force.
The thermal stochastic term $\textbf{F}_{i}^T$ in Eq.~(\ref{Md})
obeys the following conditions:
\begin{eqnarray}
\langle F_i^T(t)\rangle =0
\end{eqnarray}
and
\begin{eqnarray}
\langle F_i^T(t) F_i^T(t^\prime)\rangle =2\eta k_B
T\delta_{ij}\delta(t-t^\prime).
\end{eqnarray}

We consider a two-dimensional (2D) square simulation region $L_{x} \times L_{y}$
in the $xy$-plane
and apply periodic boundary conditions in the $x$ and $y$ directions.
For the interaction force given by Eq.~(\ref{eq-interaction_1}),
which exponentially decays for large distances, we use a cut-off for $r>8$.
The length of the square cell $L_{x} = L_{y} = L$ has been varied from $60$ to $180$
to examine the finite-size effects, and we set $L=120$ to optimize the calculation
speed without influencing the results.
To obtain stable particle patterns, we performed simulated annealing simulations (SAS)
of interacting particles.
For this purpose, particles were initially randomly distributed inside the simulation
region at some suitable non-zero temperature (depending on the inter-particle interaction).
Then temperature was gradually reduced to zero, and the simulation was continued
until a stable state was reached.
%
%After first $2000$ MD steps, the temperature was reduced to zero during next $1000$ MD steps.
%The stable state is obtained when the maximum force act on a single particle
%is smaller than $0.001$.

\begin{figure}
\vspace*{0.5cm}
\vspace*{0.5cm}
\begin{center}
\includegraphics[width=0.9\columnwidth]{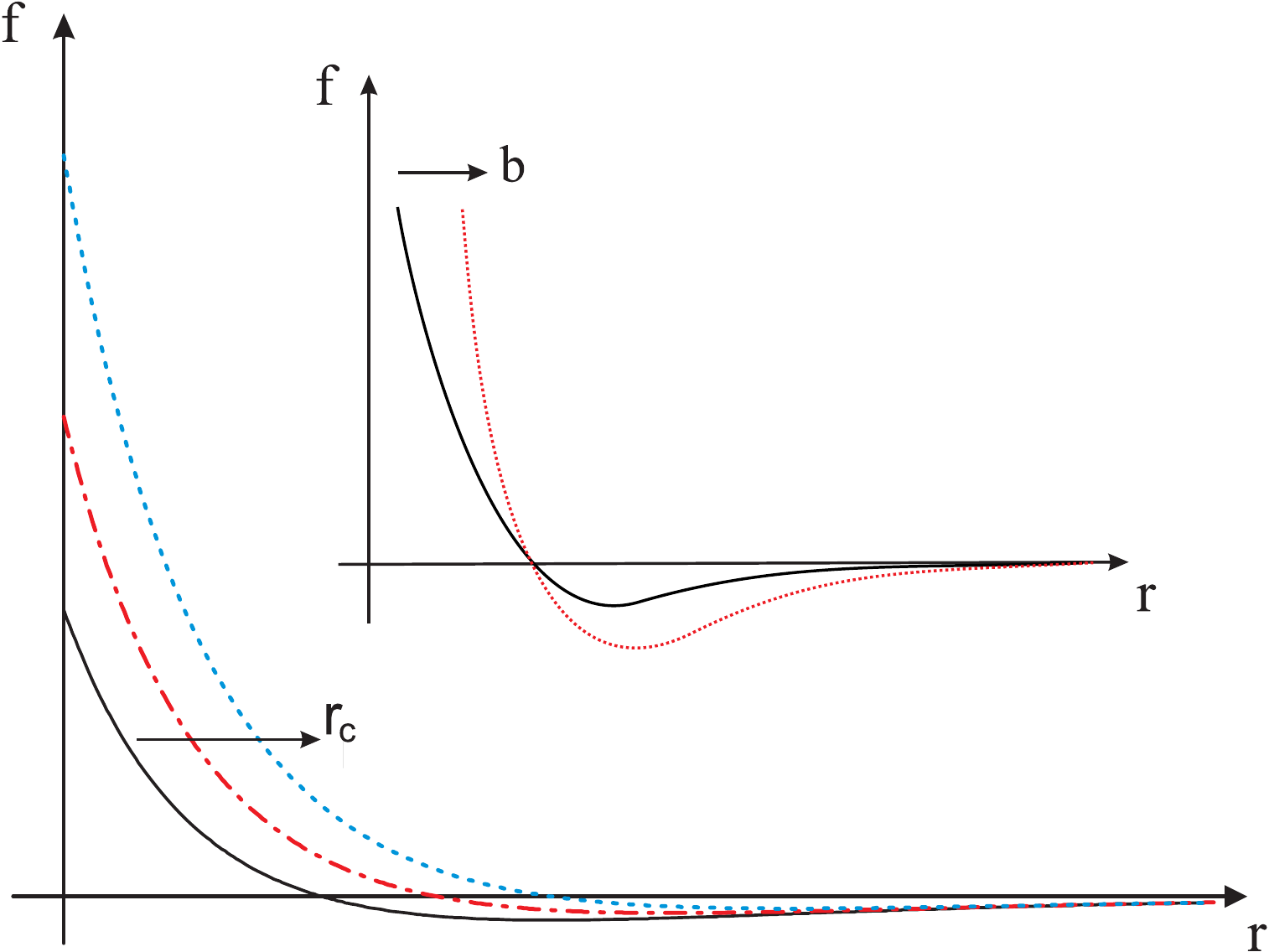}\\
\caption{
(Color online)
The profile of the inter-particle interaction force versus
the distance.
The main panel and the inset panel show the change in the force
profile due to an increase of $r_c$ and $b$, respectively. }
\label{fig-force_profile}
\end{center}
\end{figure}

\section{Patterns and phases}

In this section we will study pattern formation and identify different phases
depending on parameters of the interaction $r_c$, $b$ and the particle density $n$.
In Fig.~\ref{fig-force_profile}, we illustrate the change of the interaction
force profile due to the increase of $r_c$ and $b$.
On the one hand, the interaction force is very sensitive to $r_c$,
since its increase directly leads to the increase of the repulsive part in
Eq.~(\ref{eq-interaction_1}) (see the main panel of Fig.~\ref{fig-force_profile}).
On the other hand, increase of $b$ leads to the increase of both the repulsive
and attractive parts (see the inset of Fig.~\ref{fig-force_profile}).
However, the repulsive interaction increases much faster for $r<r_c$ than
%the increase of
the attractive interaction for $r>r_c$.

Thus, with increasing $b$ it becomes hard to
reduce the inter-particle distance below $r_{c}$.
%shrink the inter-particle distance for $r<r_c$.
In other words,
the increase of $b$ results in hardening of the core in the
interaction force (\ref{eq-interaction_1}), i.e., the interaction changes continuously from
a ``soft-core'' to a ``hard-core'' interaction.
%The increase of $r_c$ lead to the increase of the coefficient $a$ in
%Eq.~(\ref{eq-V_interaction_1}) (see eq. \ref{eq-coefficent_a}).
%Thus, the repulsive part of the interaction increase when the $r_c$ increase.

\subsection{Soft-core interaction}

\begin{figure}
\includegraphics[width=0.9\columnwidth]{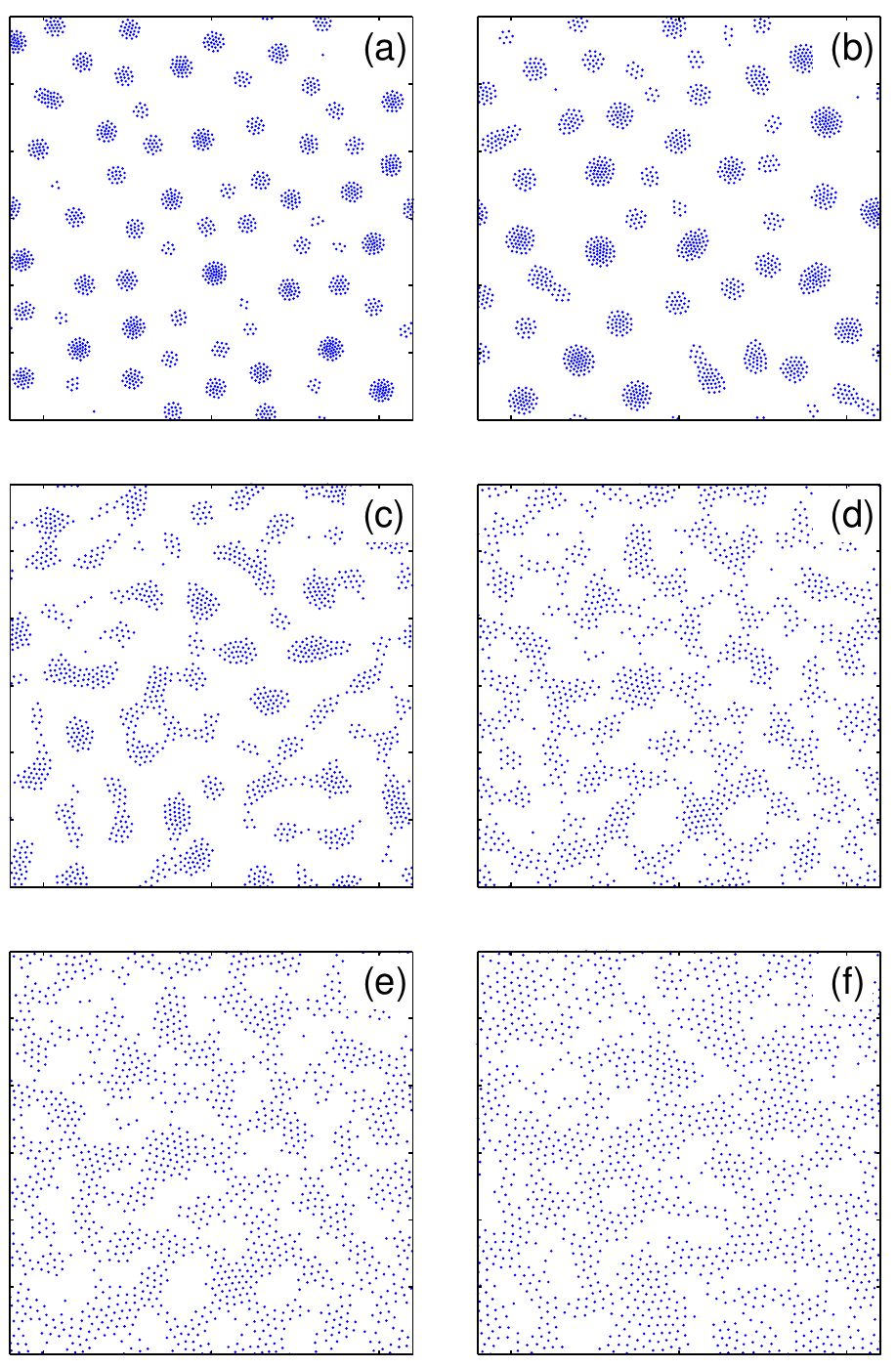}\\
\caption{
(Color online)
Different patterns for $N=1500$ particles in a unit cell $L\times L$ with $L=120$ for varying $r_c$:
$r_c=1.9$ (a), $2.1$ (b), $2.3$ (c), $2.5$ (d), $2.7$ (e), $2.9$ (f).
}
\label{fig-pattern_1500}
\end{figure}

To analyze the soft-core regime, we set the coefficient $b=1.1$.
We define the critical density as $n_c=8r_c^{-2}/\sqrt{3}$ which is the density
of an ideal (hexagonal) Wigner crystal with the lattice constant $a=r_c$.
The density is defined as $n=N/S$, where $S=120\times120$ is the area
of the simulation region and $N$ is the number of particles
(further on, when analyzing different patterns, we will refer to either
the number of particles $N$ in the simulation cell or to the density which
is $n=N/14400$).
For the case of $n<n_c$, which is considered here, the Wigner crystal
is not stable due to the attractive interaction.
In Fig.~\ref{fig-pattern_1500}, we show patterns formed by $N=1500$ particles
when $r_c$ increases from $1.9$ (a) to $2.9$ (f).
Note that the condition $n<n_c$ is always fulfilled for all the values
of $r_c$ in this range.
We found that for $r_c<2.1$ particles form clusters similar to the formation
of the ``clump'' phase found in Ref.~\cite{Reichhardt2003,PhysRevLett.90.026401,Reichhardt2005}
(see Figs.~\ref{fig-pattern_1500}(a) and (b)).
The main difference to the patterns found in Ref.~\cite{Reichhardt2003,PhysRevLett.90.026401,Reichhardt2005} is
that the relatively softer core in our case is compressed due to the attractive
interaction, and the clusters acquire a circular shape.
In addition, the interaction between the clusters (decaying exponentially
for long distances) becomes negligible for inter-cluster distance of the order
of few to several $r_{0}$.
Therefore, the clusters can be considered as non-interacting for low densities
(although still they do not approach each other),
contrary to the situation of Ref.~\cite{Reichhardt2003,PhysRevLett.90.026401,Reichhardt2005} when a super-lattice
is formed due to the long-range cluster-cluster repulsion.
When $r_c$ increases, the clusters expand.
In particular, for $r_c> 2.1$, the clusters start to elongate which is an
indication of the instability with respect to the transition to the stripe phase.
For $2.1<r_c<2.3$, a mixed state with both stripes and clusters is observed
(see Fig.~\ref{fig-pattern_1500}(c)).
Further increase of $r_c$ gradually destroys the cluster phase and leads to the
formation of the labyrinth phase (see Figs.~\ref{fig-pattern_1500}(d) to (f)).

\begin{figure}
\includegraphics[width=0.9\columnwidth]{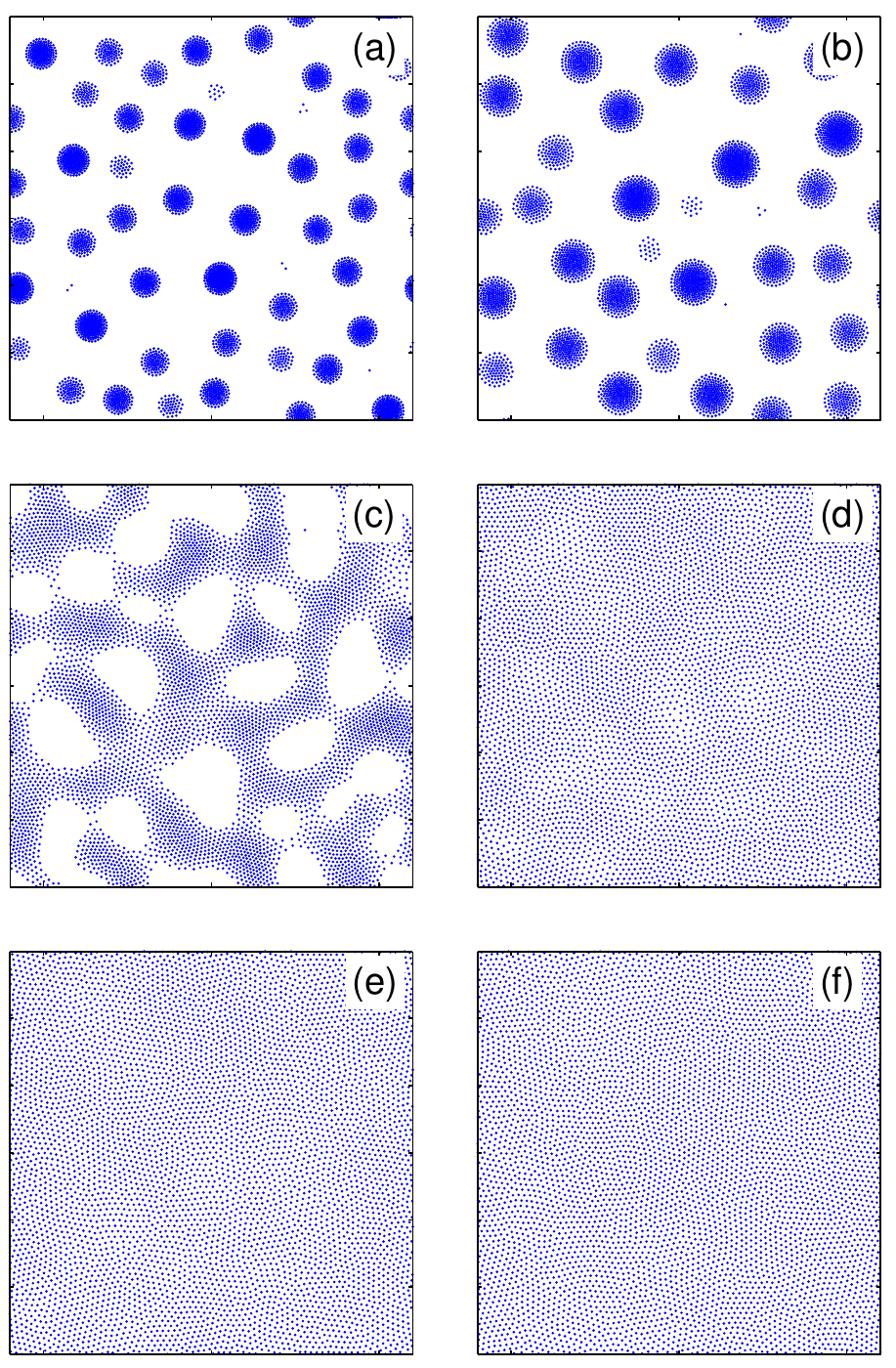}\\
\caption{
(Color online)
Patterns for $N=5500$ particles in a unit cell $L\times L$ with $L=120$ for varying $r_c$:
$r_c=1.9$ (a), $2.1$ (b), $2.3$ (c), $2.5$ (d), $2.7$ (e), $2.9$ (f).
%Patterns for number of particles $N=5500$. (a)~(e) is corresponding to %different $r_c$ as in Fig. \ref{fig-pattern_1500}.
}
\label{fig-pattern_5500}
\end{figure}

In order to investigate the influence of the density, we gradually increase
the number of particles from $1500$ to $10500$ in our simulation cell.
First, in Fig.~\ref{fig-pattern_5500}, we present patterns formed by $N=5500$
particles for varying $r_{c}$.
Note that $n>n_c$ for $N=5500$.
As compared to the lower density case shown in Fig.~\ref{fig-pattern_1500},
in the cluster phase, the additional particles lead to the expansion of the
clusters (see Figs.~\ref{fig-pattern_5500}(a) and (b)).
A mixture of clusters and stripes are formed when the additional particles form
bridges connecting the clusters (see Fig.~\ref{fig-pattern_5500}(c)) which will
be discussed in detail below.
For the labyrinth phase (shown in Figs.~\ref{fig-pattern_1500}(d) to (f) for $N=1500$),
the additional particles fill the empty regions (voids), resulting in the formation
of the triangular lattice with varying local density (Fig.~\ref{fig-pattern_5500}(d))
and, finally, a regular triangular lattice (see Figs.~\ref{fig-pattern_5500}(e) and (f)).
Note that the lattice with varying local density (Fig.~\ref{fig-pattern_5500}(d))
is not stable in systems with pure repulsive interaction such as vortices in
type-II superconductors.

\begin{figure}
\includegraphics[width=0.9\columnwidth]{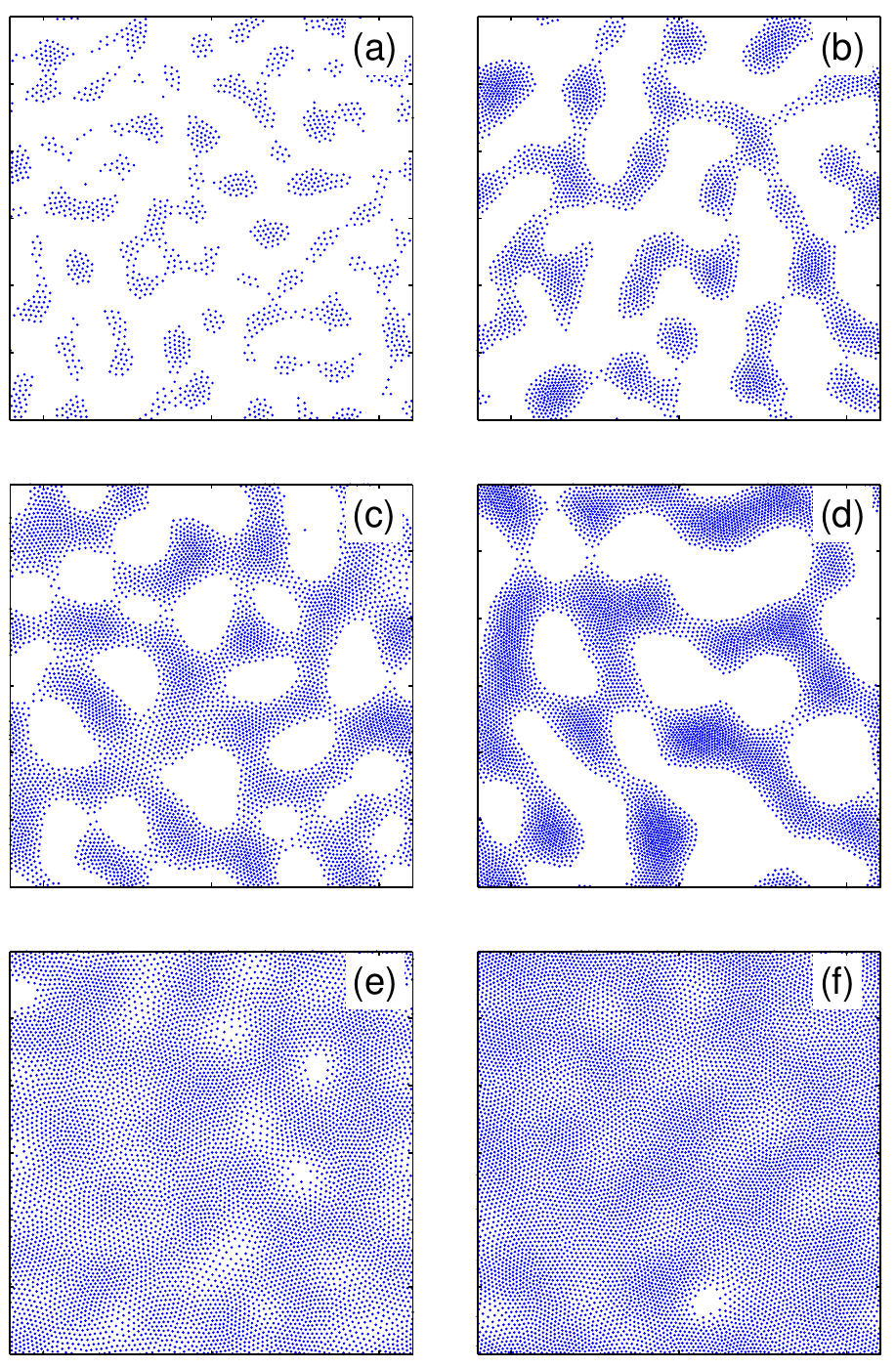}\\
\caption{
(Color online)
Patterns for $r_c=2.3$ and varying number of particles $N$
(in the computational unit cell $L\times L$ with $L=120$):
$N=1500$ (a),  3500 (b), 5500 (c),  6500 (d), 7500 (e), 9500 (f).
}
\label{fig-rc2.3}
\end{figure}

We analyzed in detail the intermediate regime (i.e., corresponding to the transition
from clusters to stripes, see Fig.~\ref{fig-pattern_1500}(c)) where $r_c\approx2.3$.
The patterns for varying density are shown in Fig.~\ref{fig-rc2.3}.
Fig.~\ref{fig-rc2.3}(a) displays a configuration at low density with $N=1500$
when many individual clusters are formed.
When increasing the density (see Fig.~\ref{fig-rc2.3}(b)), the clusters connect
with each other and form a stripe phase.
A further increase of the number of particles (Fig.~\ref{fig-rc2.3}(c)) results
in the formation of a mixed phase of interconnected stripes with voids.
A very interesting and counter-intuitive evolution is observed when the number
of particles increases from $N=5500$ to $N=6500$ (see Fig.~\ref{fig-rc2.3}(d)):
in contrast to the gradual transition from a void-rich configuration to
a lattice-rich configuration when $N$ changes from $N=3500$(b) to $N=5500$(c),
in case of $N=6500$ we observe a ``reentrant'' behavior, i.e., the void-rich phase
starts to recover which is compensated by an increase of the local density in the
stripes.
However, further increasing density results in the expansion of the stripes to
the empty regions which is accompanied by a decrease of the local density in
the stripes, as shown in Fig.~\ref{fig-rc2.3}(e).
The distribution of particles becomes more uniform, with only few small voids.
Finally, for $N=9500$ (Fig.~\ref{fig-rc2.3}(f)), we obtain a deformed triangular
lattice characterized by a varying local density with only one small void.

Our calculations show that the obtained phases are very sensitive to variations
in $r_c$.
Thus, if $r_c$ slightly decreases (e.g., $r_c=2.25$), the number of clusters
greatly increases as compared to the case $r_c=2.3$, for the same density of
particles.
For $r_c=2.25$, we also observe the transition from a void-rich configuration
to a lattice-rich configuration.
However, since the decrease of $r_c$ increases the attractive component of the
inter-particle interaction this occurs at much higher density.
For even smaller values of $r_c$, i.e., $r_c<2.1$, we do not observe the lattice
phase, even for extremely large number of particles (up to $N=20000$).

\subsection{Phase Diagram}

\begin{figure*}
\includegraphics[width=1.8\columnwidth]{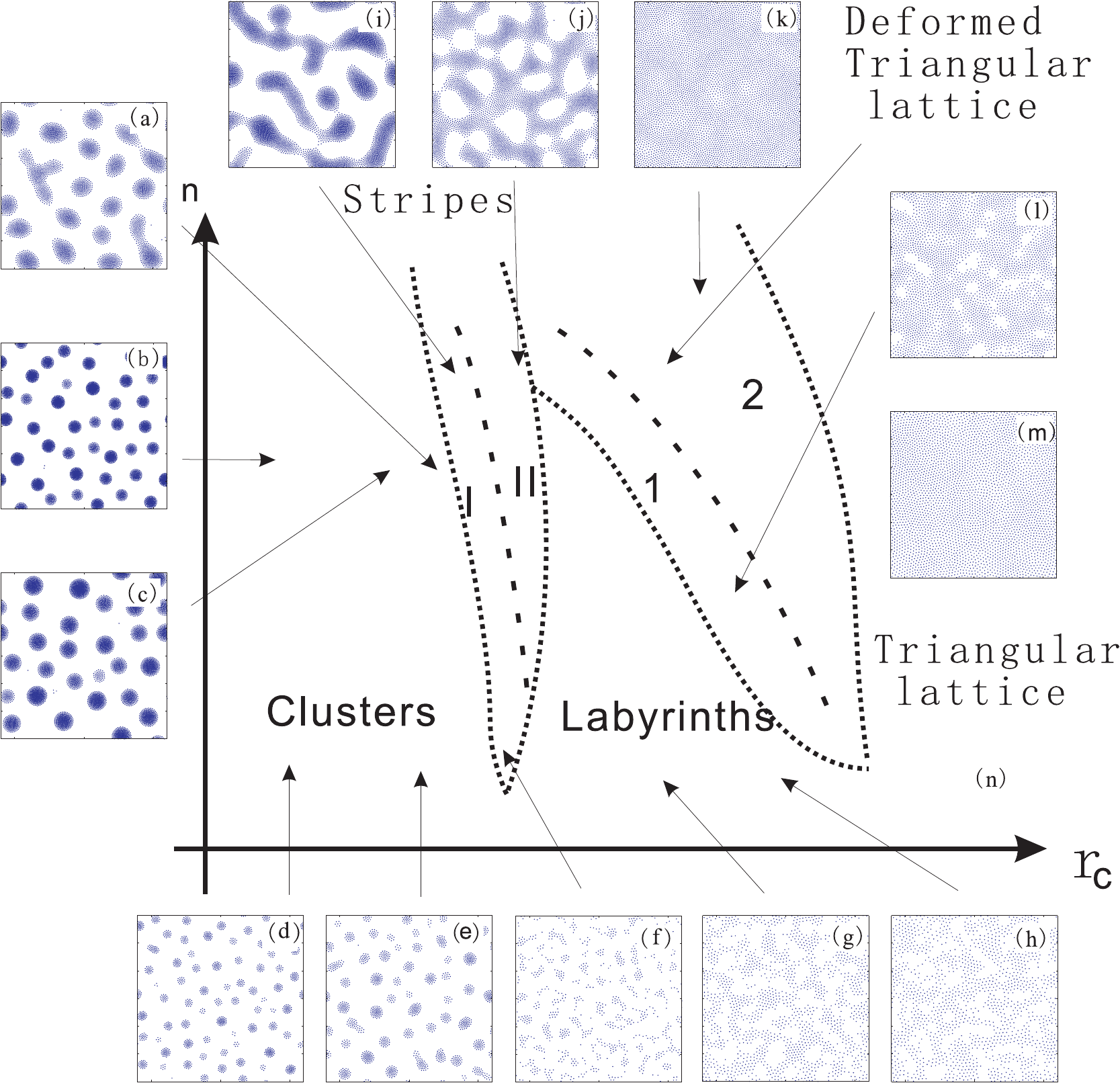}\\
\caption{
(Color online)
The phase diagram
in the plane ``critical radius $r_{c}$ $-$ density $n$'' (n)
and representative patterns (a to m) for the different phases.
For extremely low density ($n<0.03$), particles form small
well separated clusters.
For $r_c<2.1$, particles form clusters.
For larger $r_c$, the size of the clusters increases
(see the change from (d) to (e) and  (b) to (c)).
The increase of the density results in increasing the size of
the clusters (see the change from (d) to (b) and (e) to (c) ).
Further increase of $r_c$ leads to elongation of the clusters
and the
formation of ``bridges'' between them (a).
Thus, the configurations change gradually from the cluster phase
to the stripe phase (i).
When $r_c$ or $n$ become even larger, stripes interconnect
and form patterns with voids (j).
Thus the stripe phase is divided into two sub-phases, I and II.
When $r_c$ become even larger, labyrinth-like configurations are formed for $n<n_c$ (see (g) and (f)).
With further increase of $r_c$ or $n$, deformed triangular lattice
with (k) or without voids (l) are formed.
%The stripe-like configuration can also change to the deformed
%triangular lattice by increasing $n$.
The deformation of a triangular lattice is reduced for even larger
values of $r_c$ and $n$ (m).
}
\label{fig-phase}
\end{figure*}

Based on the above analysis of the patterns and phases, we constructed
a phase diagram in the plane of $r_c$ and the number density $n$
(see Fig.~\ref{fig-phase}).
For extremely low density ($n<0.03$), particles form small
clusters which are well separated, and the patterns are rather
insensitive to variations of $r_c$.
However, phases become richer when density $n$ increases (i.e., $n>0.1$).
Low values of $r_c$ (i.e., $r_c<2.1$) still favor cluster formation for a
broad range of densities although the clusters become denser for large $n$.
With increasing $r_c$, clusters become unstable with respect to the formation
of ``bridges'' between separate clusters which is a precursor of the formation
of stripes.
Stripes are formed in a rather narrow range of $r_c$ when $r_c>2.1$
(see Fig.~\ref{fig-phase}).
The stripe phase is represented by two sub-phases, I and II,
i.e., void-rich phase (see Fig.~\ref{fig-phase}(i))
and void-poor (lattice-rich) phase (Fig.~\ref{fig-phase}(j)).
For larger $r_c$ and $n<n_c$, particles form labyrinth structures.
However, when increasing the density, additional particles fill the
empty regions and finally they form a deformed triangular lattice
with varying density.
Depending on $r_c$ and $n<n_{c}$ deformed lattice is characterized by
a varying local density or by the appearance of voids.
Correspondingly, we distinguish two regions (1 and 2 in Fig.~\ref{fig-phase}).
Note that in the vicinity of the phase boundaries patterns are
always mixtures of the two phases (e.g., clusters and stripes)
except the phase transition from the stripe phase to the deformed
triangular lattice.
Near this phase boundary, particles either form stripes with high
local density or deformed triangular lattices with lower local density.
The probability of finding the lattice configuration greatly increases
when $r_c$ or $n$ increases.
Finally, for very large density, when the average interparticle distance
becomes larger than $r_c$ the repulsive interaction prevails resulting
in the formation of an almost perfect triangular lattice.

The phases presented in Fig.~\ref{fig-phase} are found in many
real systems.
For example, the cluster phase is found in such systems as
colloids~\cite{PhysRevLett.97.078301,Sciortino2005,PhysRevLett.94.208301,PhysRevLett.93.055701} and neutron
stars~\cite{PhysRevC.69.045804}.
The obtained stripe patters are very similar to those in Langmuir
monolayers~\cite{Suresh1988}.

Several of the calculated phases were found in superconductors,
e.g., the stripe phase in the mixed state of type-I superconductors,
clusters of Meissner phase or vortex clusters in the intermediate
state in low-$\kappa$ type-II superconductors~\cite{Brandt1987,Brandt2011},
the labyrinth phase (i.e., vortex lattice with voids) or vortex
clusters in type-1.5 superconductors~\cite{PhysRevLett.102.117001}.
In spite of the variety of the physical systems and length scales
ranging from nano- and micro-objects to cosmic objects, the common
feature of all these systems is a competing attractive-repulsive
interparticle interaction which allows analyzing them within the
same approach.

%%%%%%%%%%%%% I am here %%%%%%%%%%%%%

\subsection{Hard-core interaction}

Let us now analyze the influence of coefficient $b$ on the pattern formation.
As mentioned above (see Fig.~\ref{fig-force_profile}), an increase
in $b$ changes the potential from a soft-core to a hard-core.
Hardening the repulsive core in the interaction potential generally
leads to a decreasing compressibility of the inner parts of the patterns
where particles are closely packed.
Correspondingly, the patterns change as compared to the soft-core
regime.

\begin{figure}
\includegraphics[width=0.9\columnwidth]{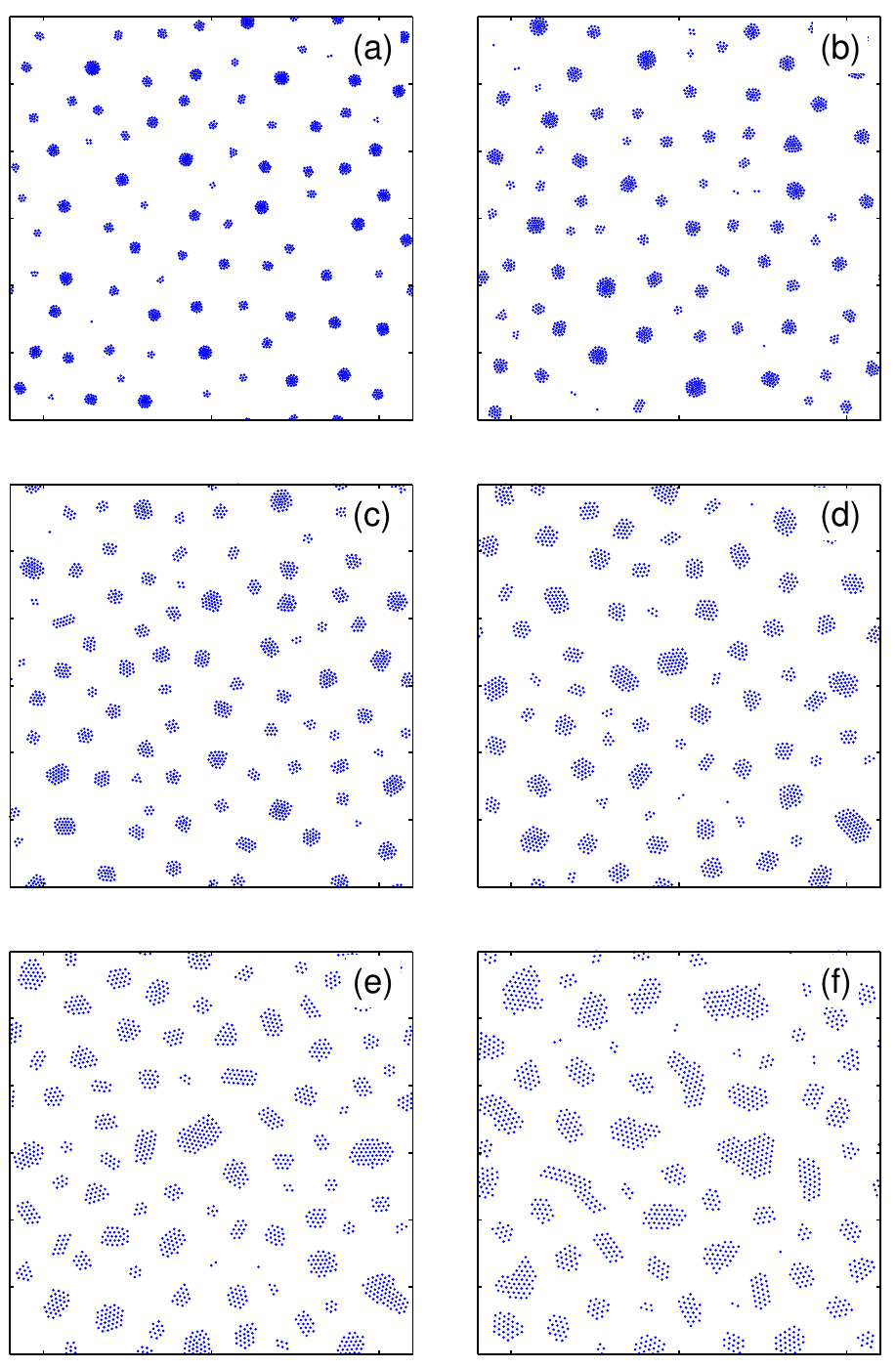}\\
\caption{
(Color online)
Patterns for $N=1500$ particles in a unit cell $L\times L$ with $L=120$, $b=4$ and for varying $r_c$:
$r_c=1.15$ (a), $1.3$ (b), $1.45$ (c), $1.6$ (d), $1.75$ (e), $1.9$ (f).
}
\label{fig-pattern_1500_2}
\end{figure}

As shown in Fig.~\ref{fig-pattern_1500_2}, for $b=4$ and low density
($N=1500$), all the patterns are clusters of different shape.
Thus for $r_c=1.15$ and $r_c=1.3$, the clusters are of circular shape
similar to those found in the soft-core regime
although for $r_c=1.3$ some clusters composed of a small number
of particles have polygon shapes.
With increasing $r_c$, polygon shaped clusters become more favorable,
which allows us to identify them as short stripes.
For even larger $r_c$, the clusters become much larger.
They represent separate islands of triangular lattices.

\begin{figure}
\includegraphics[width=0.9\columnwidth]{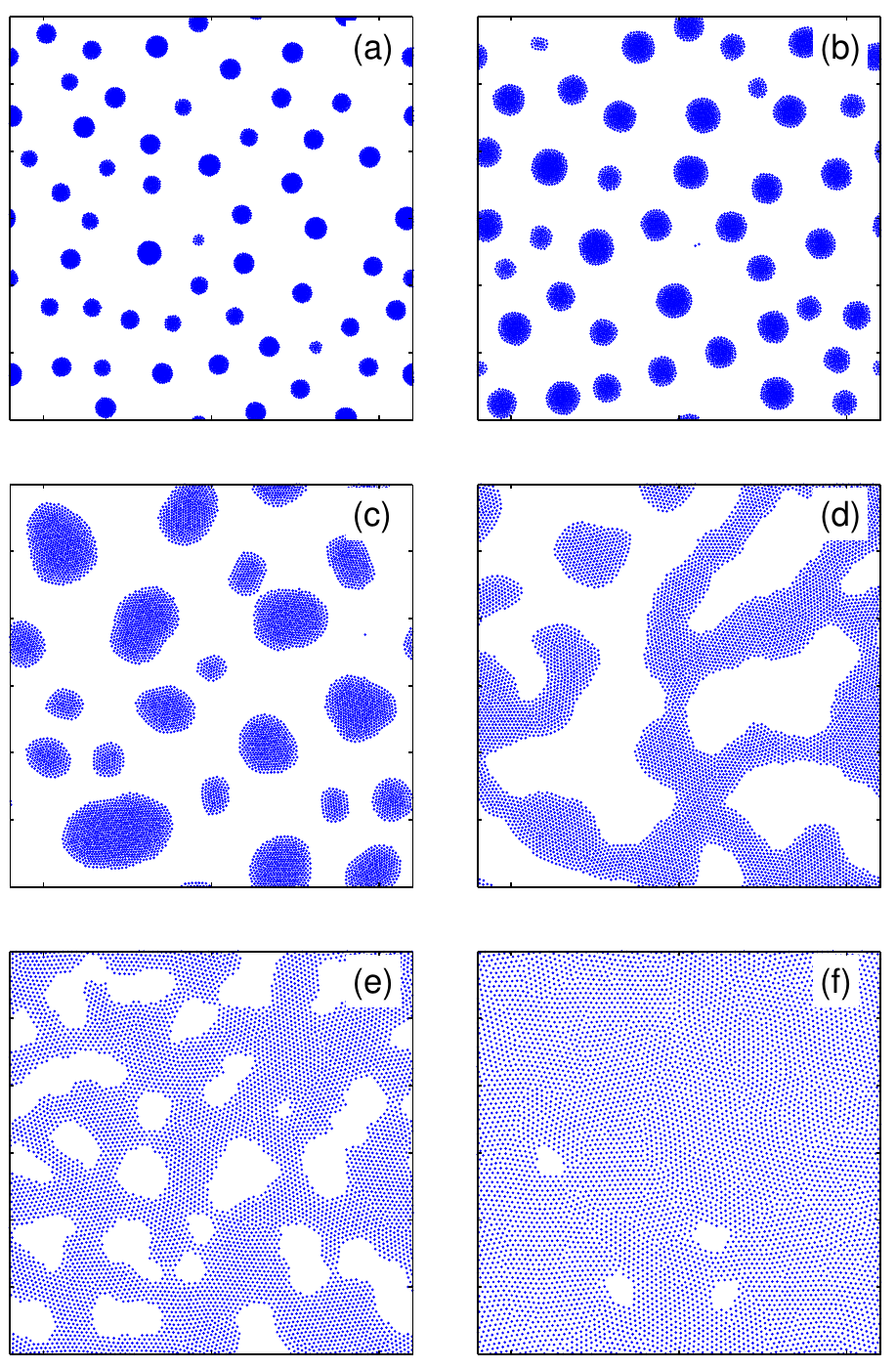}\\
\caption{
(Color online)
Patterns for $N=6500$ particles in a unit cell $L\times L$ with $L=120$, $b=4$ and for varying $r_c$:
$r_c=1.15$ (a), $1.3$ (b), $1.45$ (c), $1.6$ (d), $1.75$ (e), $1.9$ (f).
}
\label{fig-pattern_6500_2}
\end{figure}

For higher density ($N=6500$), the variety of phases is much richer.
Although for $r_c=1.15$ circular-shape clusters are still observed
(see Fig.~\ref{fig-pattern_6500_2}(a)),
which grow in size for $r_c=1.3$ and slightly deviate in shape
from circular (Fig.~\ref{fig-pattern_6500_2}(b)),
for larger $r_c=1.45$ deviations of the cluster shape from circular
become more pronounced (see Fig.~\ref{fig-pattern_6500_2}(c)).
With further increasing $r_c$, i.e., $r_c=1.6$, stripes are formed
(see Fig.~\ref{fig-pattern_6500_2}(d)) followed by
lattices with voids for $r_c=1.75$ and $r_c=1.9$.
In the hard core regime, deformations of lattices occur via the
appearance of voids instead of varying local density
(in the soft-core regime).

\section{Analysis of the patterns}

\subsection{Radial Distribution Function}

Although different patterns and phases studied above are qualitatively
well distinguished, it is highly desirable to build a set of solid
criteria which would allow us to identify them in a quantitative manner.
For this purpose, we analyze here different phases by means of the
Radial Distribution Function (RDF).
The RDF, $g(r)$, describes the variation of the atomic (particle) density
as a function of the distance from one particular atom (particle).
If we define an average density as $n = N/V$
(where $V$ is the volume or surface area in the 2D case),
then the local density at
distance $r$ is $ng(r)$.
The knowledge of RDF is important since one can measure $g(r)$ experimentally
using neutron scattering or x-ray diffraction scattering~\cite{Hajdu1976}.
Moreover, macroscopic thermodynamic quantities can be calculated using $g(r)$
in thermodynamics \cite{Frenkel2002}.

\begin{figure}
\vspace*{0.5cm}
\vspace*{0.5cm}
\includegraphics[width=1\columnwidth]{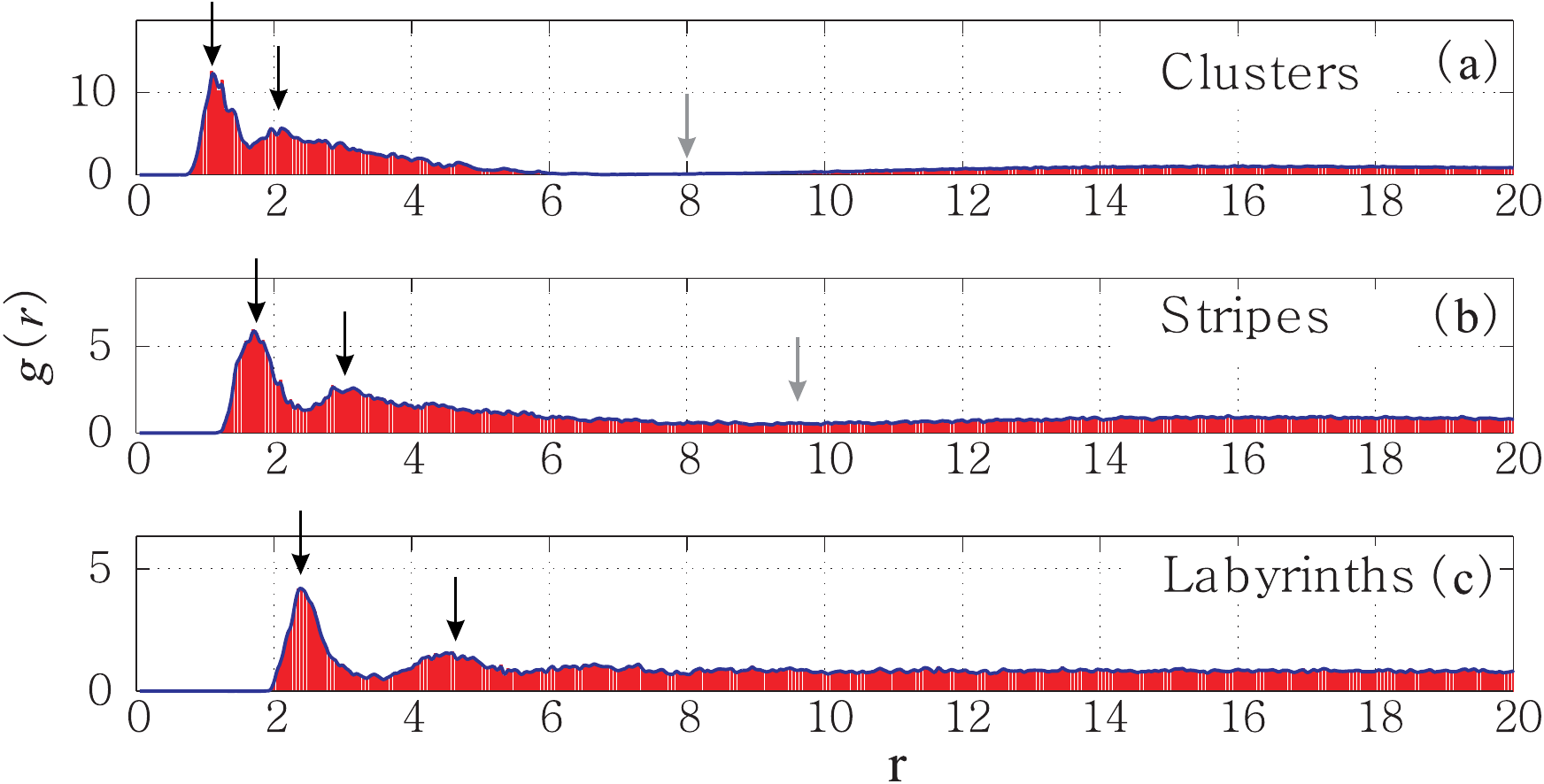}\\
\caption{
(Color online)
The average RDF of the patterns formed at low density ($N=1500$).
(a), (b), and (c) correspond to the patterns shown in
Figs.~\ref{fig-pattern_1500}(b), (d), and (f), respectively.
The peaks marked by dark arrows show the shot-range periodicity.
The gray arrow shows the minimum of the RDF.
}
\label{fig-rdf1}
\end{figure}

In our calculations, we define the RDF $g_i(r)$ as follows:
\begin{equation}
\label{eq-rdf}
 g_i(r)=\frac{\triangle N/\triangle r}{2\pi r}.
\end{equation}
Here, the lower index indicates that the RDF centers at the position of the
$i$th particle; $\triangle N$ is the number of particles whose distance to the $i$th
particles is between $r$ and $r+\triangle r$.
The average $g(r)$ is given by
\begin{equation}
\label{eq-rdfa}
g(r)=\frac{1}{N}\sum_{i=1}^{N}g_i(r).
\end{equation}
%
%The different phases can be analyzed by calculating RDF $g(r)$.
In Fig.~\ref{fig-rdf1}(a), we plot the function $g(r)$ calculated
for the low-density cluster configuration shown in
Fig.~\ref{fig-pattern_1500}(b).
The function $g(r)$ has two well-pronounced peaks.
The first peak corresponds to the average distance to the first
coordination sphere (nearest neighbors) while the second one is
located at the distance approximately twice the distance to the
first peak, which shows short-range periodicity.
For $r$ larger than the size of the cluster, and smaller than the
inter-cluster distance, $g(r)<1$, since only few particles are
located inside this range.
As well as for the RDF obtained for the pasta phase in neutron
stars\cite{PhysRevC.69.045804}, the decreasing tail in our case also shows a
strong aggregation of the particles in the cluster phases.
The minimum of $g(r)$ is very close to zero since most of the
clusters have circular symmetry and they are well separated.
The position of the minimum of $g(r)$ (marked by the gray arrow in
Fig.~\ref{fig-rdf1}(a)) gives an estimate of the average diameter
of the clusters.

For the stripe phase (see Fig.~\ref{fig-rdf1}(b) for the pattern
shown in Fig.~\ref{fig-pattern_1500}(d)), there are also two peaks
indicating the short-range periodicity, similar to the cluster phase.
However, unlike in the case of clusters the minimum of $g(r)$ is
non-zero since the stripes are generally not separated.
For the labyrinth phase (see Fig.~\ref{fig-rdf1}(c) for the pattern
shown in Fig. \ref{fig-pattern_1500}(f)), only short-range periodic
ordering exists, and $g(r)$ becomes uniform for larger $r$.

\begin{figure}
\vspace*{0.5cm}
\includegraphics[width=1\columnwidth]{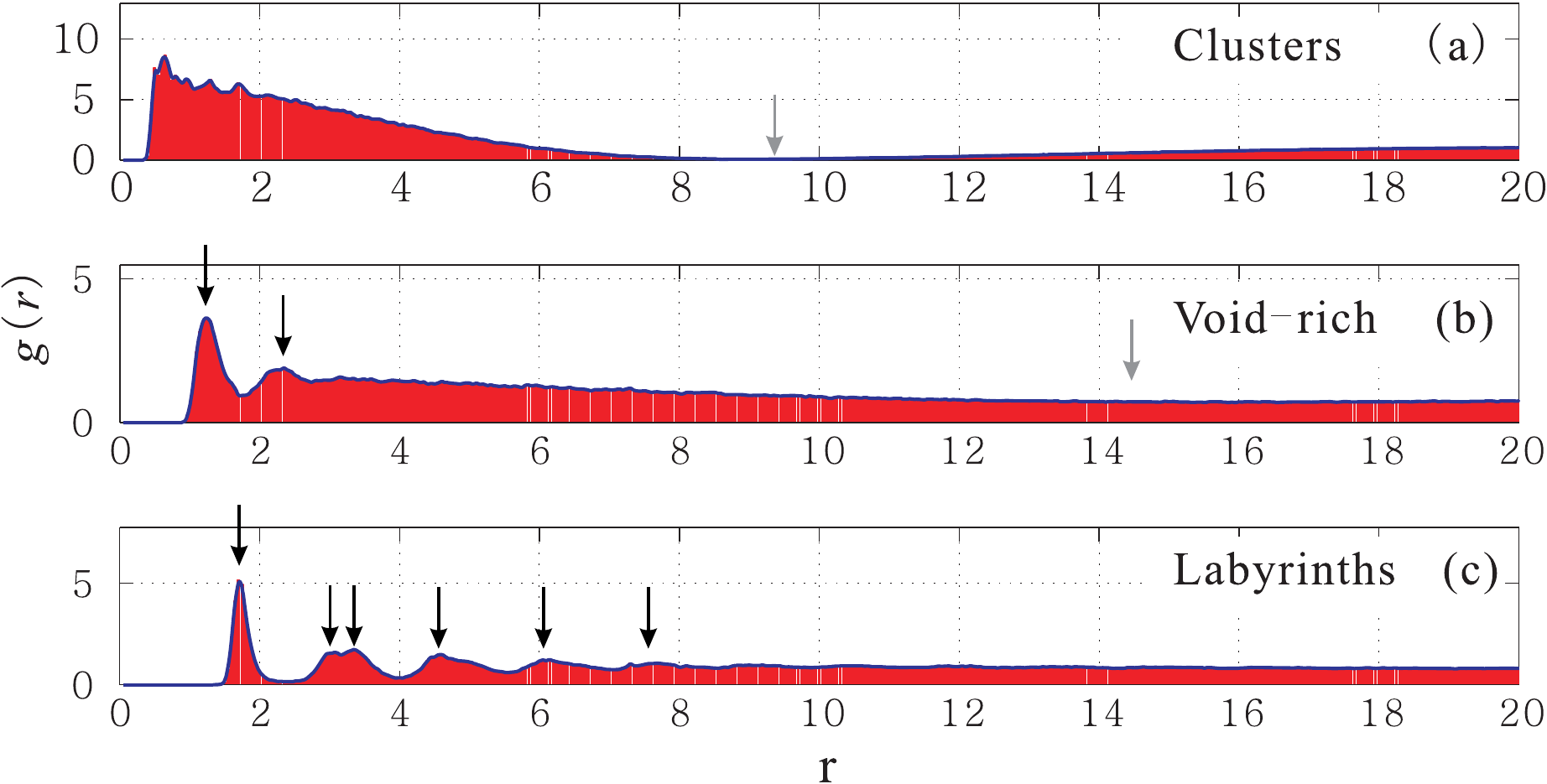}\\
\caption{
(Color online)
The average RDF of the patterns formed at high density (N=5500).
(a), (b), and (c) correspond to the patterns shown in
Figs.~\ref{fig-pattern_5500}(b), (d), and (f), respectively.
The arrows have the same meaning as in Fig.~\ref{fig-rdf1}.
}
\label{fig-rdf2}
\end{figure}

In Fig.~\ref{fig-rdf2}, we plot $g(r)$ for the patterns formed at
high density (N=5500).
For clusters, the increase in the density leads to an increasing
variation of the cluster size.
Note, however, that the sizes of individual clusters are rather
insensitive to moderate variations of the number of particles in the
clusters, due to the strong compressibility of the core.
As a result, the function $g(r)$ for the high density clusters misses
the short-range periodicity and the second peak disappears
(see Fig.~\ref{fig-rdf2}(a)).
Strikingly, this peak still exist for the void-rich phase
(see Fig.~\ref{fig-rdf2}(b)).
The decrease of the tail of $g(r)$ becomes very slow, which shows
a minor aggregation of the particles.

For the lattice phase (see Fig.~\ref{fig-rdf2}(b)), the position of
the first peak is $r_1\approx a=\sqrt{2/(\sqrt{3}n)}$, where $a$ is
the distance between two neighboring particles in the ideal triangular
lattice with the density $n$.
The positions of the second and third peaks are at $r_2\approx \sqrt{3}a$
and  at $r_2\approx 2a$, respectively, corresponding to those in
a triangular lattice.
A fourth, fifth and even sixth peaks appear, which shows that this
phase is much more ordered.
However, due to the variation of the local density, these peaks are
strongly broadened, and they are actually a combination of many neighboring
peaks.
These peaks become clearer for larger $r_c$ or $n$, which implies that
the lattices become more regular.

It is interesting to compare the results of the function
$g(r)$ for hard-core interaction with those for soft-core interaction.
Although the clusters shown in Figs.~\ref{fig-pattern_6500_2}(a) and (b)
have shapes very similar to those of the clusters shown in
Figs.~\ref{fig-pattern_5500}(a) and (b),
the analysis using the RDF shows that most of the clusters shown
in Fig.~\ref{fig-pattern_6500_2}(b) have hexagonal ordered cores.
The peaks in $g(r)$ appear to be much better separated than for, e.g.,
the deformed triangular lattice formed in the soft-core case
(see Fig.~\ref{fig-rdf2}(c)).
For the stripe and void-rich phases, the RDF shows much clearer
triangular lattice ordering.

Therefore, the analysis using the RDF allowed us to establish
quantitative criteria for the different phases and reveal the differences
in the structure of the patterns (which look similar, e.g., clusters)
in case of soft- and hard-core interparticle interaction.

\begin{figure}
\includegraphics[width=1\columnwidth]{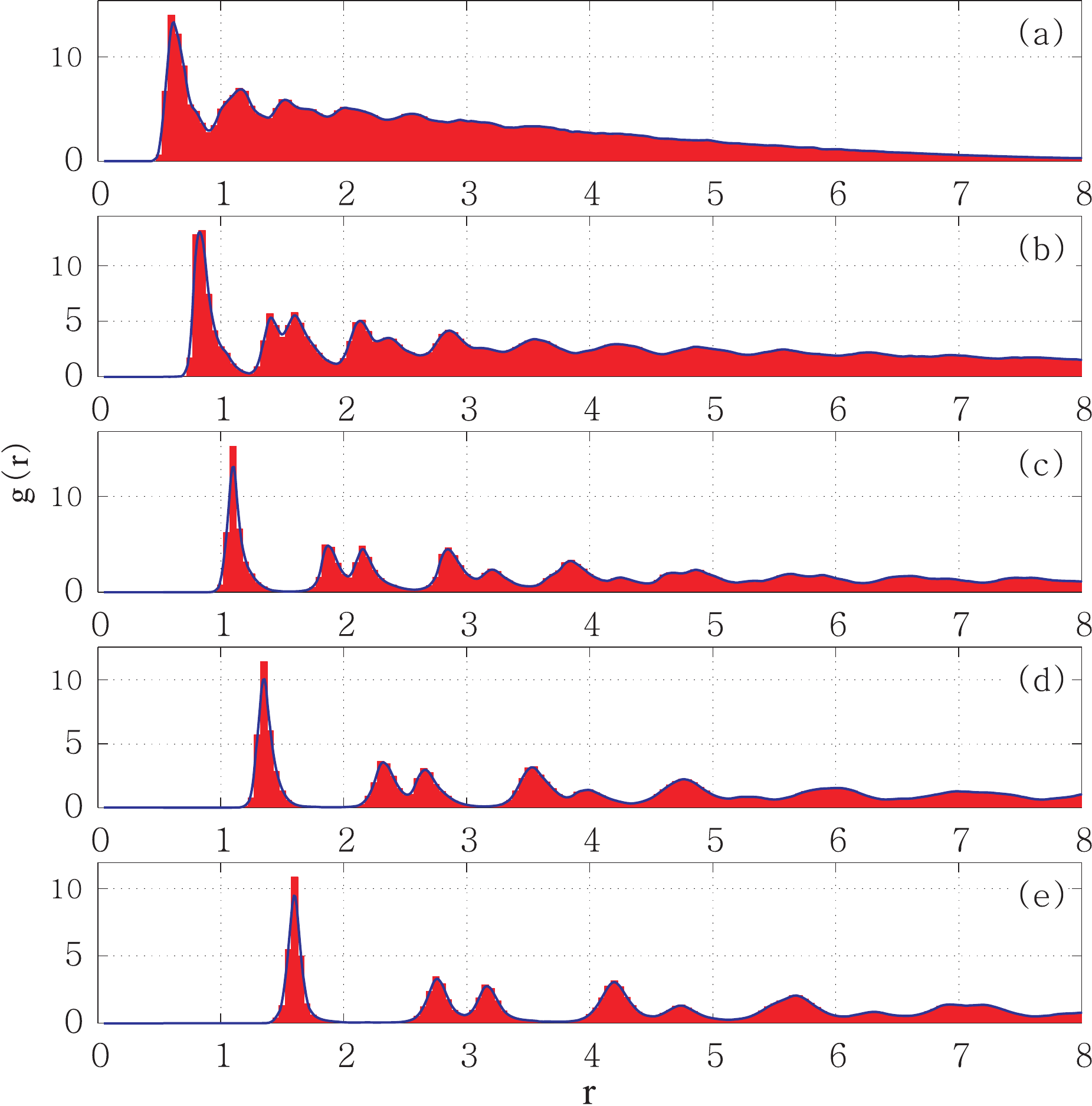}\\
\caption{
(Color online)
The average RDF of the patterns formed in the hard-core case.
(a), (b), (c), (d), and (e) correspond to the patterns (a), (b), (c), (d), and (e) in Fig.~\ref{fig-pattern_6500_2}, respectively.
}
\label{fig-rdf_h}
\end{figure}

\subsection{Local density}

\begin{figure}
\includegraphics[width=0.9\columnwidth]{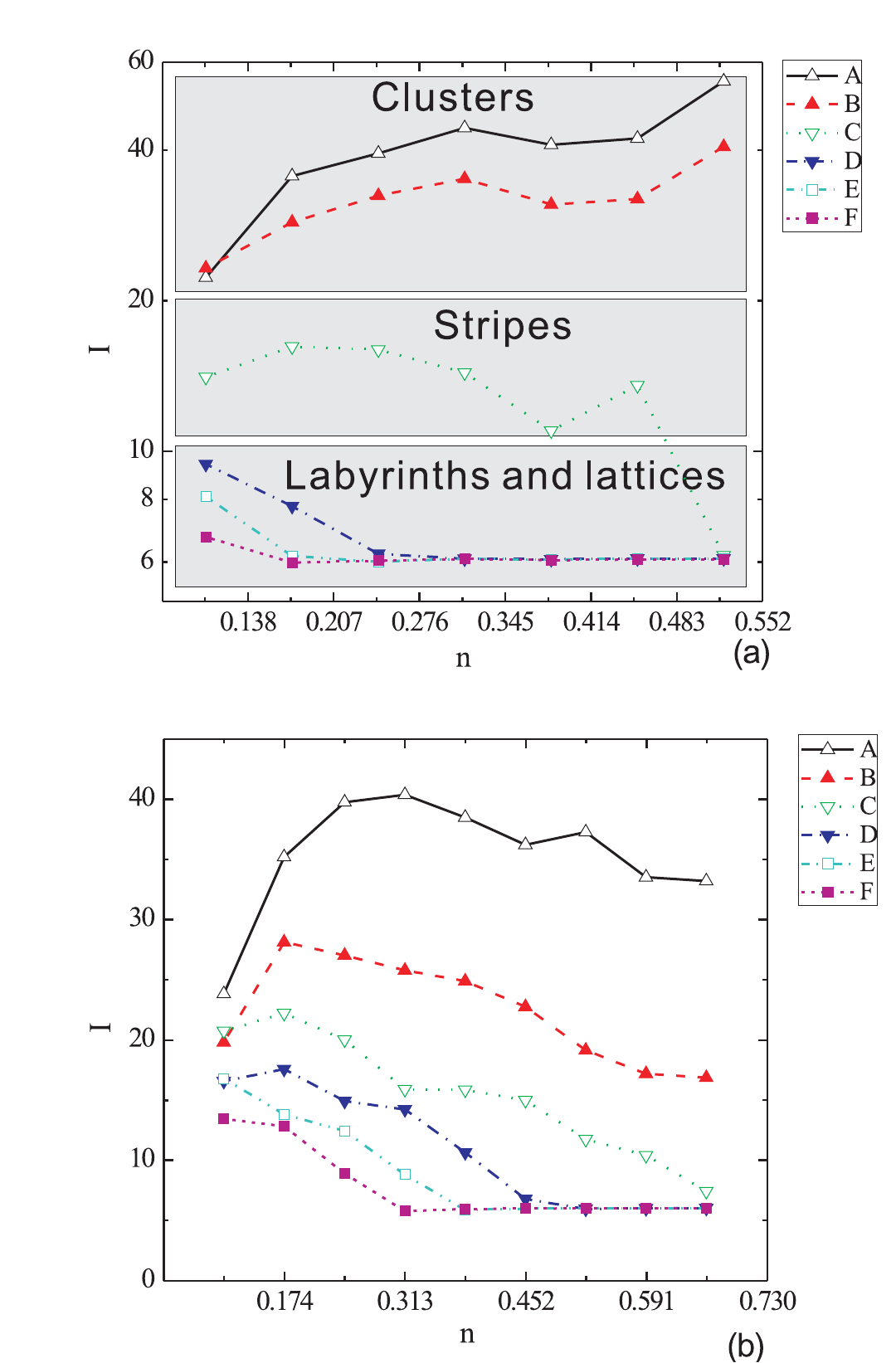}\\
\caption{
(Color online)
The average local density $I$ versus the particle density $n$ for $b=1.1$ (a).
Curves A, B, C, D, E, and F correspond to
$r_c=1.9$, $2.1$, $2.3$, $2.5$, $2.7$, and $2.9$, respectively.
For the cluster phase, $I > 20$,
for the stripe phase, $10 < I < 20$,
for the labyrinth phase, $6 < I < 10$.
$I=6$ for lattice and deformed lattice configurations.
The local density $I$ versus the number of particles $N$ for $b=4$ (b).
Curves A, B, C, D, E, and F correspond to
$r_c=1.15$, $1.3$, $1.45$, $1.6$, $1.75$, and $1.9$, respectively.
}
\label{fig-order_I}
\end{figure}

Let us define, using the RDF, an ``order parameter'' to characterize
the different phases.
By definition, the local density is:
\begin{equation}
\label{eq-rdf_integral}
 I_i=\int_0^\xi 2\pi r n g_i(r)dr= N_\xi-1.
\end{equation}
Here, $N_\xi\approx \pi \xi^2 n$ is the average number of particles
within the circle centered at the $i$th particle with radius $R=\xi$.
Since in an ideal triangular lattice one particle has six nearest
neighbors, we take $N_\xi=7$, then
$\xi=\sqrt{N_\xi/\pi n}=\sqrt{7/\pi n}$.
Thus, for any configuration characterized by a small local density
fluctuation, the average local density $I = \langle I_i \rangle = 6$.
From the definition given by Eq.~(\ref{eq-rdf_integral}), we can see
that the presence of a large fraction of empty regions can considerably
increase
$I$ since only the ``shell'' of those regions of thickness $\xi$
is considered.
In Fig.~\ref{fig-order_I}(a), we plot the function $I$ for different
$r_c$ and $N$ for the soft-core interaction with $b=1.1$.
We see that the function $I$ can be used to charaterize the different
phases.
Thus we found that, for clusters, $I$ is always large ($I>20$)
since aggregation is strong.
For stripes, aggregation is smaller, and thus $I$ is smaller
($10<I<17$).
For labyrinths, the regions free of particles are relatively small.
Therefore, $I$ ranges between $6$ to $9$.
Finally, for the lattice phase, $I$ is always $6$.
Therefore, the function $I$ serves as a measure of the aggregation of
particles and allows us to effectively distinguish the different phases.

The function $I$ also provides a tool to analyze the stability of
patterns with increasing density.
This is demonstrated in Fig.~\ref{fig-order_I}(b), where we plot
$I$ for the hard-core case when $b=4$.
For $r_c=1.15$, $I>20$ for all the values of the density, and thus the
clusters are stable.
However, the situation changes for $r_c=1.3$: while clusters are formed
for low densities up to $N \approx 7500$, for larger density $I$ decreases
below the value $I=20$ which means that clusters elongate
and interconnect giving rise to the onset of the stripe phase.
For  $r_c=1.45$, the particles start to form stripes at even lower
density (N=3500).
For higher density, the additional particles fill in the empty
regions and finally they form the lattice phase.
For $r_c>1.6$, the island-like clusters formed at low densities are
very unstable with respect to an increase in density.
The additional particles rapidly occupy the empty regions, and the
patterns change form clusters to stripes and from stipes to lattice
with increasing density.

Therefore, the phenomenological description of pattern evolution with
increasing density we revealed above in Sec.~III.A has been verified in
a quantitative manner using the rigid basis of the local density function
approach.

\subsection{Occupation factor}

\begin{figure}
\includegraphics[width=0.9\columnwidth]{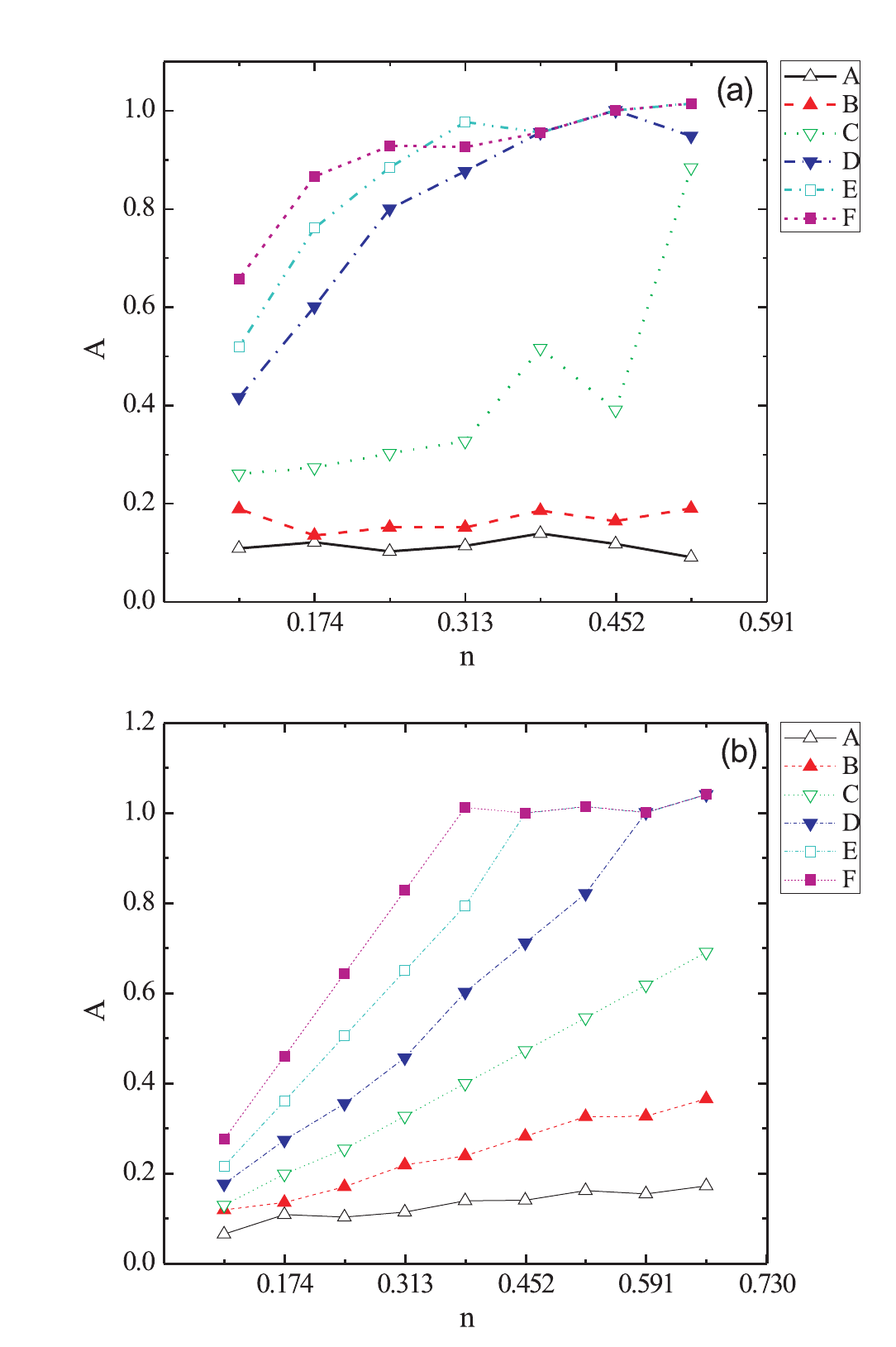}\\
\caption{
(Color online)
The occupation factor $A=(r_1/a)^2$ for $b=1.1$ (a).Curves A, B, C, D, E, and F correspond to
$r_c=1.9$, $2.1$, $2.3$, $2.5$, $2.7$, and $2.9$, respectively.
The occupation factor, $A$ for $b=4$ (b).
Curves A, B, C, D, E, and F correspond to
$r_c=1.15$, $1.3$, $1.45$, $1.6$, $1.75$, and $1.9$, respectively.
}
\label{fig-order_II}
\end{figure}

We demonstrated that the (average) RDF $g(r)$ (Eq.~(\ref{eq-rdfa}))
and the local density function $I$ (Eq.~(\ref{eq-rdf_integral}))
allowed us to unambiguously characterize the different patterns
and phases.
At the same time, it is still hard to distinguish, using the above
tools, between the labyrinth phase (or a lattice with voids) and
the lattice phase (especially in the soft-core regime when the
corresponding RDF does not differ much from each other, and both phases
are characterised by small values of $I$.
However, these phases can be easily distinguished by employing another
simple criterion, namely, the occupation factor which characterizes
the fraction of the space occupied by particles to the total space.

Let us assume that all the patterns are formed by islands of triangular
lattice with the average distance between two nearest neighbor
particles $r=r_1$, where $r_1$ is the position of the first peak
of the RDF.
Then the ratio of the area occupied by the particles and the whole
simulation area is $A=(r_1/a)^2$.
As shown in Fig.~\ref{fig-order_II}(a), this ratio can be used to
distinguish the labyrinth phase and the lattice phase, since for
the lattice phase $A\approx 1$.

We also found that circular clusters in the case of soft-core
interaction are very stable with respect to an increase in the density.
In this case, the additional particles cannot efficiently increase
the occupation factor due to the increase in the local density of
the core.
However, for the stripe and labyrinth phases, the occupation factor
$A$ increases with the density (see Fig.~\ref{fig-order_II}(a)).
These phases are not stable for high density.
Thus the large jump in $A(N)$ at $N=6500$ in curve C in
Fig.~\ref{fig-order_II}(a) shows the phase transition from the stripe
phase to the lattice phase.

As one can expect, in the case of hard-core interaction, the occupation
factor increases nearly linearly versus the density (see Fig.~\ref{fig-order_II}(b)).
Therefore, the polygon shape and island-like clusters are not stable:
with increasing density they evolve to the lattice phase (the plateau
in $A(N)$ in Fig.~\ref{fig-order_II}(b)).

%The above analysis provided us with a tool to identify the clusters
%and stripes.
In addition, let us introduce another useful quantity which
characterizes the degree of ``perfection'' of a lattice.
Let us define the particles with the inter-particle distance shorter
than $\xi$ as neighboring particles.
Then
 \begin{align}\label{eq-rdf_displacement}
 \varepsilon&=\frac{1}{a}\sqrt{\frac{\int_0^\xi 2\pi r g(r) n (r-a)^2 dr}{\int_0^\xi 2\pi r g(r) n dr}}\nonumber\\
  &=\frac{1}{a}\sqrt{\frac{\int_0^\xi  r g(r)  (r-a)^2 dr}{\int_0^\xi  r g(r) dr}}.
\end{align}
is an average displacement of the inter-distance between two neighbor
particles (measured in units of $a$), which is independent of the
density.
The function $\varepsilon$ is non-zero for a deformed triangular
lattice and $0$ for the ideal triangular lattice.
Thus, $\varepsilon$ quantitatively measures the degree of perfection
of a lattice.
Note that $\varepsilon$ is only used for lattices as an auxiliary tool
to distinguish ordered lattices from less ordered ones (see Table I).

\section{Conclusions}

Using molecular-dynamics simulations,
we analyzed the pattern formation and identified different phases
in a system of particles interacting via a non-monotonic potential,
with a repulsive short-range part and attractive long-range part.
The form of the interacting potential is generic: it describes,
depending on specific parameters, the interparticle interaction
in a variety of physical systems ranging from, e.g., atoms and
molecules (Lennard-Jones potential) to colloids and neutron stars.
It can also be used as a model of inter-vortex interaction in
low $\kappa$ type-II superconductors and in recently discovered
so-called ``type-1.5'' superconductors.
The obtained different phases were summarized in a phase diagram
in the plane ``critical radius $r_{c}$ $-$ density $n$'' ($r_{c}$
is the critical radius where the interaction force changes its sign).
We also analyzed the influence of the hardness of the ``core'', i.e.,
the strength of the repulsive core part of the interaction potential
on the pattern formation.

We developed a set of criteria in order to unambiguously identify the
obtained phases using the following approaches:
(i) the Radial Distribution Function (RDF) $g(r)$,
(ii) the local density function $I$, and
(iii) the occupation factor $A$.
In addition, we introduced a parameter which characterizes the degree
of perfection of a lattice $\varepsilon$.
Employing these approaches, we elaborated the criteria for
the identification of the different phases which are summarized
in Table~\ref{criterion}.
%The parameter $I$ helps us find out the clusters and stripes phase.
%There is a overlap of the labyrinths phases and lattice phases,
%which can be identified with the help of parameter $A$.

%
%\begin{table}[h]
%\begin{center} {\footnotesize
%\begin{tabular}{|c|c|c|}
%\hline
%Patterns &  I  &   A\\\hline
%clusters &  $>20$ &  \\ \hline
%stripes &  $10\sim20$  & \\ \hline
%labyrinths & $6 \sim 10$ &$<1$ \\\hline
%lattices & $\approx6$ &$\approx1 $ \\\hline
%\end{tabular} }
%\end{center}
%\caption{\footnotesize .The criterion for the different phases. The
%clusters and stripes phase is well separated. The labyrinths phases
%and lattice phases overlaps in some region, which can be separated
%by the help of $A$. }
%\label{criterion}%\end{table}
%

\begin{table}[h]
\begin{center} {\footnotesize
\begin{tabular}{|c|c|c|c|c|}
\hline
Patterns   & $g(r)$    &  $I$  &   $A$  &  $\varepsilon$ \\ \hline
clusters   & peak at $r_{1}$($r_{2}$), &  $>20$      & & \\
           & $g(r)_{min}=0$            &             & & \\ \hline
stripes    & peak at $r_{1}$($r_{2}$), & $10\sim 20$ & & \\
           & $g(r)_{min}>0$            &             & & \\ \hline
labyrinths & peak at $r_{1}$($r_{2}$), & $6 \sim 10$ & $<1$ & \\
           & $g(r)\approx$ const, $r>r_{2}$ &        & & \\ \hline
lattice    & several peaks: $r_{1}$, $r_{2}$\ldots & $\approx 6$ &$\approx 1 $ & 0 \\ \hline
deformed   & several peaks: $r_{1}$, $r_{2}$\ldots & $\approx 6$ &$\approx 1 $ & $>0$ \\
lattice    &                           &             & & \\ \hline
\end{tabular} }
\end{center}
\caption{
%\footnotesize
The set of criteria used to quantitatively identify the different phases
in terms of
the RDF $g(r)$,
the local density function $I$,
the occupation factor $A$,
and the parameter $\varepsilon$ (for details, see the text).
}
\label{criterion}
\end{table}

\section{Acknowledgments}

We acknowledge fruitful discussions with Ernst Helmut Brandt.
This work was supported by the ``Odysseus'' Program of the Flemish
Government and the Flemish Science Foundation (FWO-Vl),
the Interuniversity Attraction Poles (IAP) Programme --- Belgian
State --- Belgian Science Policy, and the FWO-Vl.
%

%\bibliography{ref_2}

\end{document}